\documentclass[showpacs,twocolumn,aps,prl]{revtex4}
\usepackage{amssymb,epsfig,graphicx}
\usepackage{natbib}

\newcommand{\isittrueornot}{false}

\begin{document} 
\title{Cavitation-induced force transition\\ 
in confined viscous liquids under traction} 
\author{Sylwia Poivet, Fr\'ed\'eric Nallet, 
Cyprien Gay, Pascale Fabre}
\affiliation{Centre de recherche Paul-Pascal--CNRS, 
Av. Schweitzer, 33600 Pessac, France}
%\homepage{http://www.crpp.u-bordeaux.fr/tack.html}
\email{fabre@crpp.u-bordeaux.fr}
\date{May 31, 2002}
\pacs{82.35.Gh, 47.55.Bx, 81.70.Bt, 83.50.Jf}

\begin{abstract} 
We perform traction experiments on simple liquids highly confined
between parallel plates. At small separation rates, we observe a
simple response corresponding to a convergent Poiseuille flow.
Dramatic changes in the force response occur at high separation rates,
with the appearance of a force plateau followed by an abrupt drop. By
direct observation in the course of the experiment, we show that
cavitation accounts for these features which are reminiscent of the
utmost complex behavior of adhesive films under traction. Surprisingly
enough, this is observed here in purely viscous fluids.
\end{abstract}

\maketitle 

\vspace{\baselineskip}

Adhesive materials are often tested by way of the so-called probe-tack
test~\cite{ZOSEL} which tends to mimic the detachment of an adhesive
joint. The material is used as a thin film (typically $100\mu{\rm m}$
in thickness) deposited on a flat, rigid substrate. A flat, solid
punch (also called indenter or probe) is approached and kept in
contact with the film for a few seconds. It is then pulled away at
some prescribed, constant velocity, while the applied force is
recorded. Good adhesives typically yield a force response with the
following characteristic features~\cite{CRETONFABRE,PHYSTOD}: the
force increases sharply and reaches a peak, followed by a plateau at a
lower value, until it drops quite abruptly to much lower values and
finally vanishes when complete separation is achieved. In recent
years, direct observation of the film during traction has been
conducted~\cite{CRETON}. Meniscus
instabilities~\cite{SHULL,CHAUDHURY}, bubble growth and fibril
formation are commonly observed. Some of these phenomena have been
associated to particular features of the force response: force peak
for the bubble appearance~\cite{CRETON} and force plateau for the
bubble growth~\cite{CRETON} or fibril elongation~\cite{ZOSEL}. Such
behaviors have been attributed to the specific rheological properties
of adhesive materials.
Otherwise, many studies have been devoted to the instabilities of
viscous liquids confined between two parallel plates~\cite{HELESHAW}
under traction, a geometry sometimes named ``lifted Hele-Shaw cell''.
During the traction, the fluid has to flow inwards. Due to the
resulting pressure gradient, the edge of the sample destabilizes from
its initial, regular shape through the Saffman--Taylor
mechanism~\cite{SAFFMAN,MAHER}. As these instabilities develop, air
fingers grow towards the center of the sample, 
producing characteristic fingering patterns~\cite{TARAFDAR}. 
At the end of the traction process, 
interesting instabilities occur prior to the detachment of the liquid
column~\cite{MCKINLEY}.

In the present Letter, we study the force response of viscous liquids
in combination with pattern observations. Our aim is to determine
which phenomena observed in adhesive materials rely on their specific
properties and which of them are more general.  We show indeed that
the force plateau and subsequent drop observed in adhesives are also
present in viscous liquids and that the mechanisms involved are
similar.

\vspace{\baselineskip}

The apparatus consists in two plane, horizontal plates whose
separation can be varied. 
A drop of liquid is initially deposited on the bottom plate. The glass
top plate is slowly approached until the drop is confined into a film
of prescribed thickness. The top plate is then pulled at constant
velocity $V$ (in the range $1\mu{\rm m/s}-1{\rm mm/s}$) while the
force is being recorded {\it via} a transducer. 

Following the mechano-optical design described in
Ref.~\cite{TORDJEMAN}, 
we observe the liquid--glass interface through a built-in
internally-reflecting prism used as the top plate~\cite{WHYTOP}.
In addition, a picture is made of the patterns
left in the liquid on the lower plate after full separation is
achieved.

The liquid used is a highly viscous, non-volatile 
silicon oil (Rhodia 47V1000000,
viscosity $\eta=10^3{\rm Pa.s}$). In all experiments, the sample
volume is kept constant. Initially, the thickness is
$h_0=100\mu{\rm m}$ and the diameter of the squeezed drop is $9{\rm
mm}$. The oil has been left in a vacuum chamber (pressure
$10^{-4}\,{\rm atm}$) for one hour in order to remove entrapped gas
bubbles and achieve a reproducible initial sample state.

We plot the force versus the plate separation in a log-log
representation. The separation is obtained by subtracting the machine
elongation from the nominal separation~\cite{FRANCISHORN}. The
machine rigidity $K=7.5\,10^5{\rm N/m}$ has been measured separately.

\vspace{\baselineskip}

\begin{figure}
\center
\includegraphics[width=0.9\linewidth]{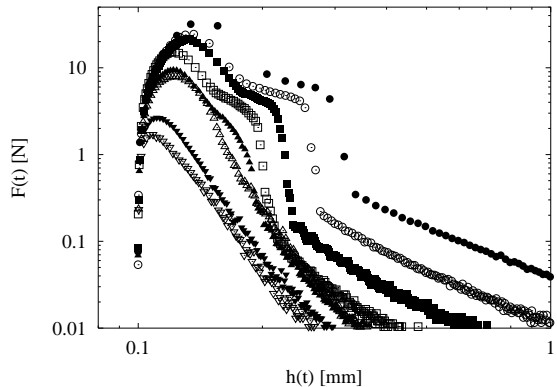}
\caption{Force versus plate displacement at separation rates ranging
from $V=1\mu{\rm m/s}$ ($\triangledown$) to $V=1{\rm mm/s}$
($\bullet$). The force plateau and drop appear only above some
critical velocity ($\blacktriangle,\,V_c=15\mu{\rm m/s}$).}
\label{tous-regimes}
\end{figure}

\begin{figure*}
\center
\includegraphics[width=0.7\linewidth,height=0.21\linewidth,draft=\isittrueornot]{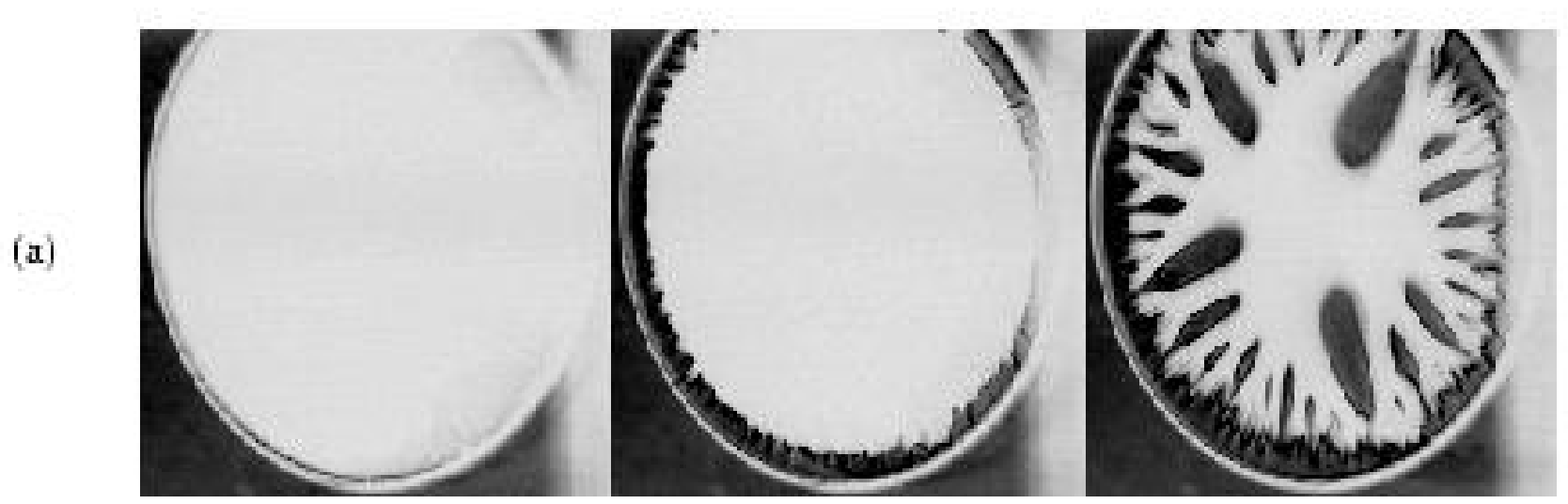}
\includegraphics[width=0.28\linewidth,draft=false]{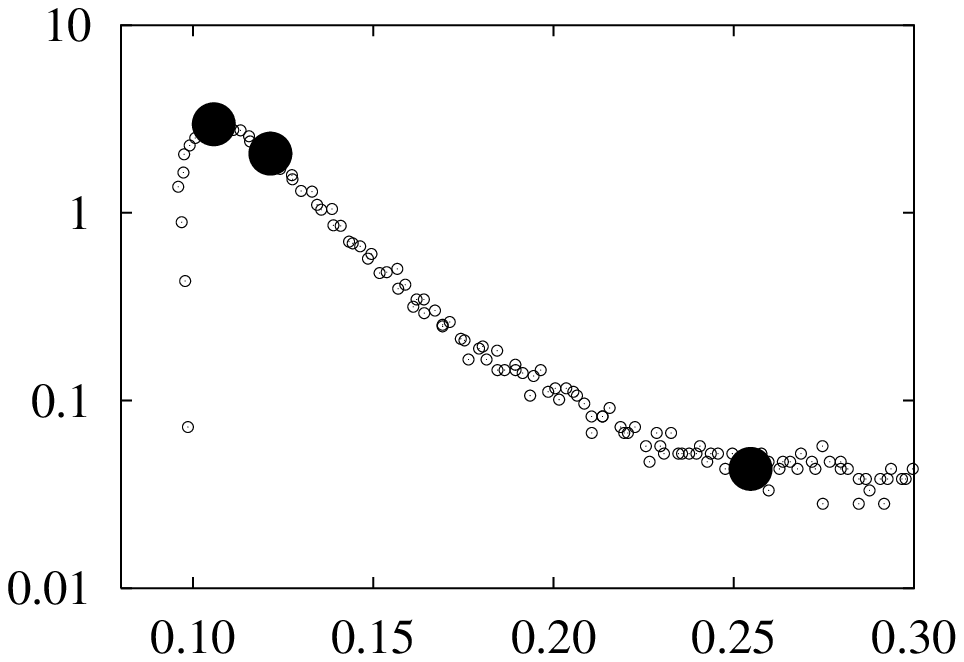}

\vspace{-1\baselineskip}

\includegraphics[width=0.7\linewidth,height=0.21\linewidth,draft=\isittrueornot]{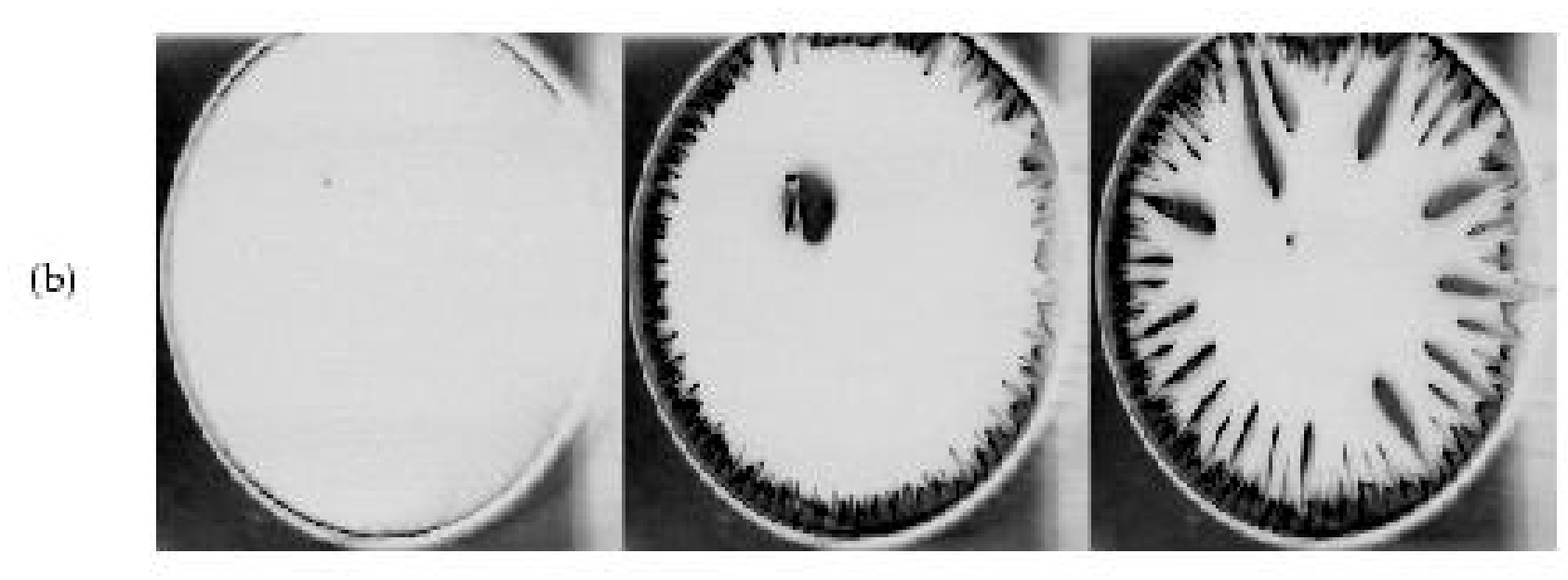}
\includegraphics[width=0.28\linewidth,draft=false]{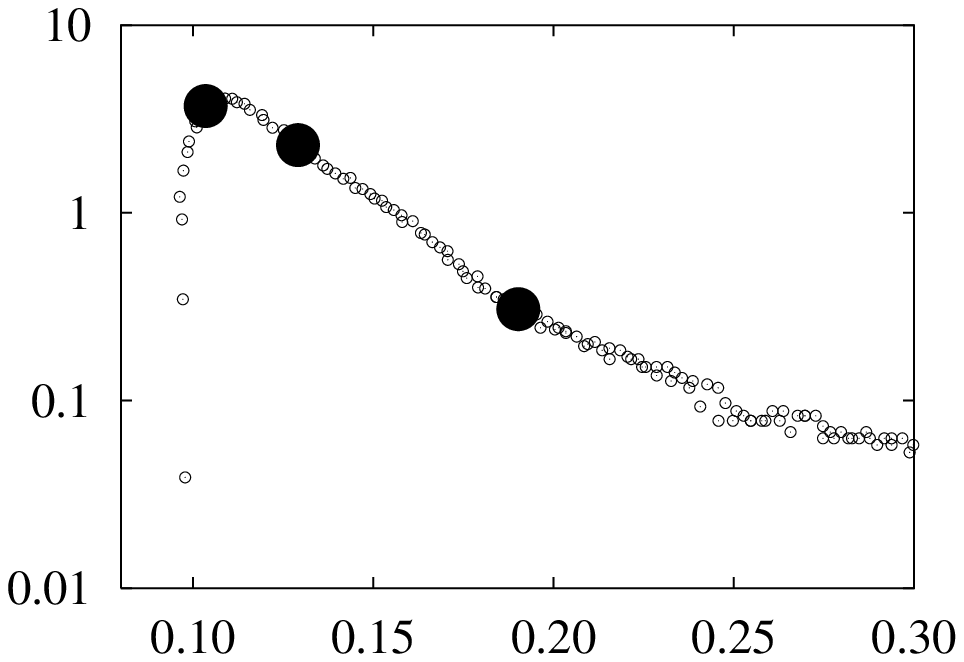}

\vspace{-.67\baselineskip}

\includegraphics[width=0.7\linewidth,height=0.21\linewidth,draft=\isittrueornot]{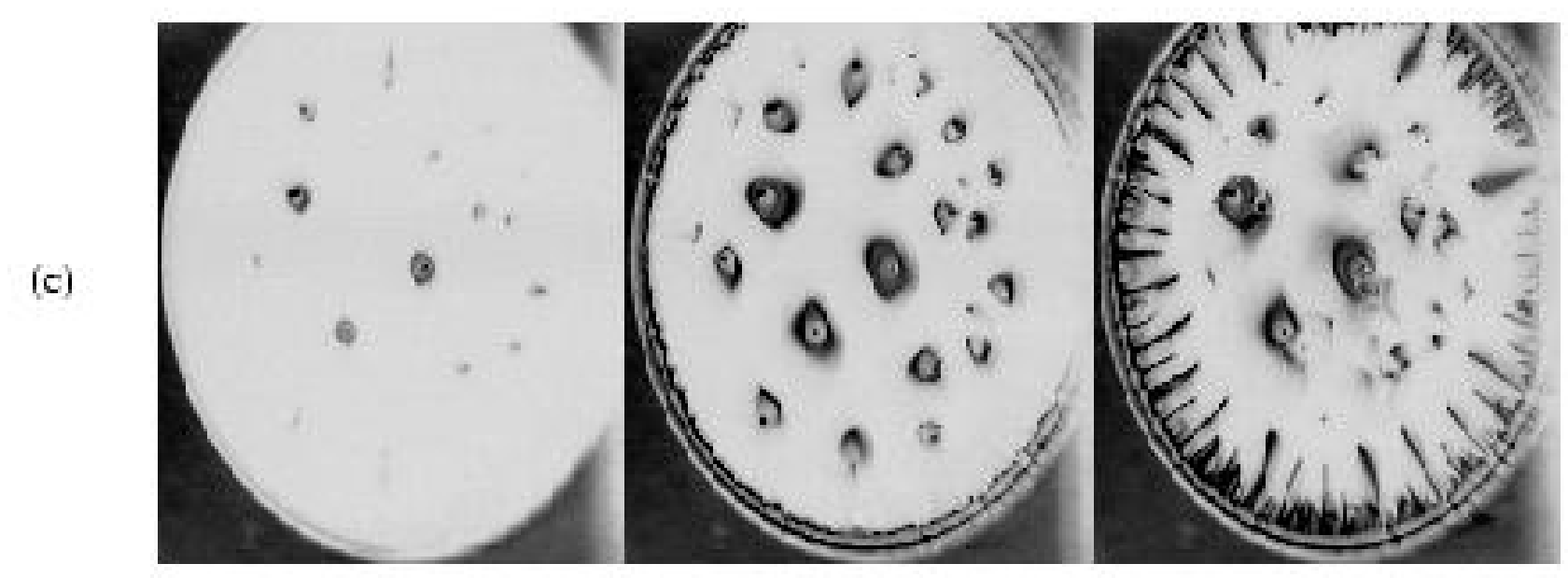}
\includegraphics[width=0.28\linewidth,draft=false]{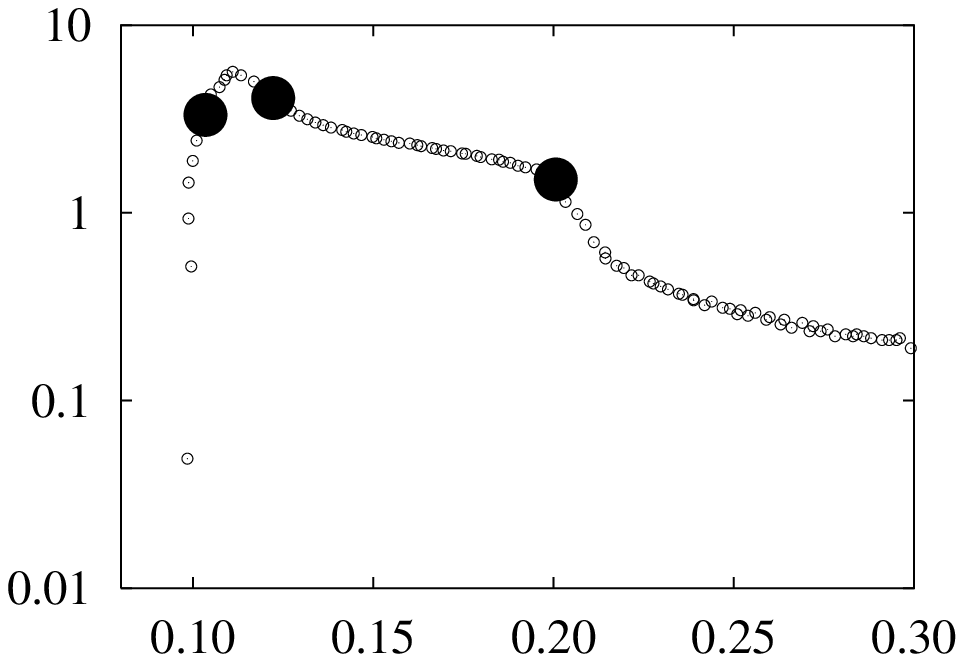}

\caption{Instantaneous pictures of the liquid--glass interface during
the traction experiment at {\bf (a)} $V=8\mu{\rm m/s}<V_c$, 
{\bf (b)} $V=16\mu{\rm m/s}\simeq V_c$ 
and {\bf (c)} $V=100\mu{\rm m/s}>V_c$, as revealed through a built-in
internally-reflecting prism used as the top plate. Inverted gray
scale: white regions are wet by the fluid and do not reflect light;
the glass surface is in contact with vapor in dark regions. The
corresponding force--separation curves are displayed, 
with filled circles indicating when the images were recorded.}
\label{bepego}
\end{figure*}

\begin{figure}
\begin{minipage}{0.33\linewidth}
\center\includegraphics[width=\linewidth,height=0.75\linewidth]{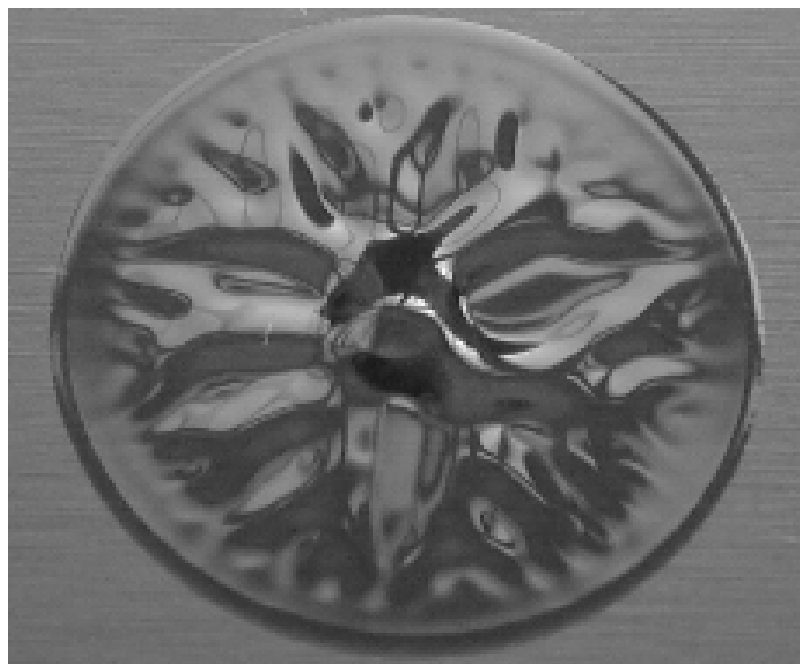}

\vspace{0.5\baselineskip}

\includegraphics[width=\linewidth,height=0.75\linewidth]{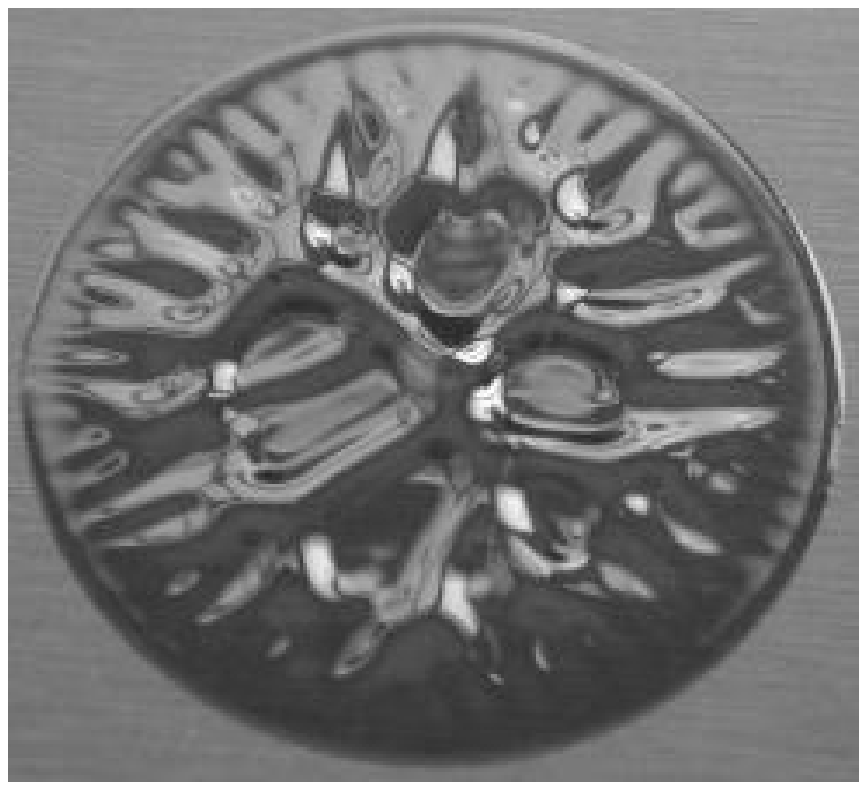}
\end{minipage}\hfill
\begin{minipage}{0.33\linewidth}
\center\includegraphics[width=\linewidth,height=0.75\linewidth]{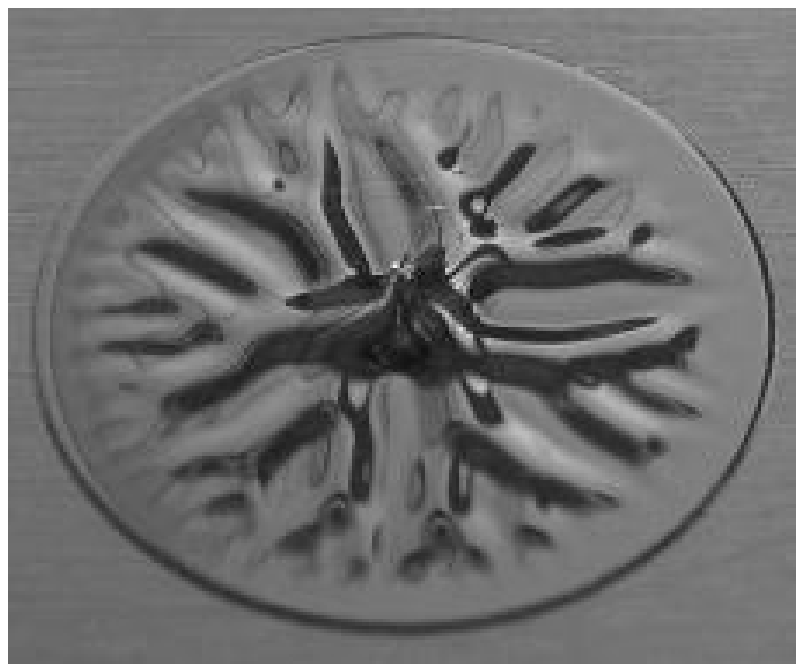}

\vspace{0.5\baselineskip}

\includegraphics[width=\linewidth,height=0.75\linewidth]{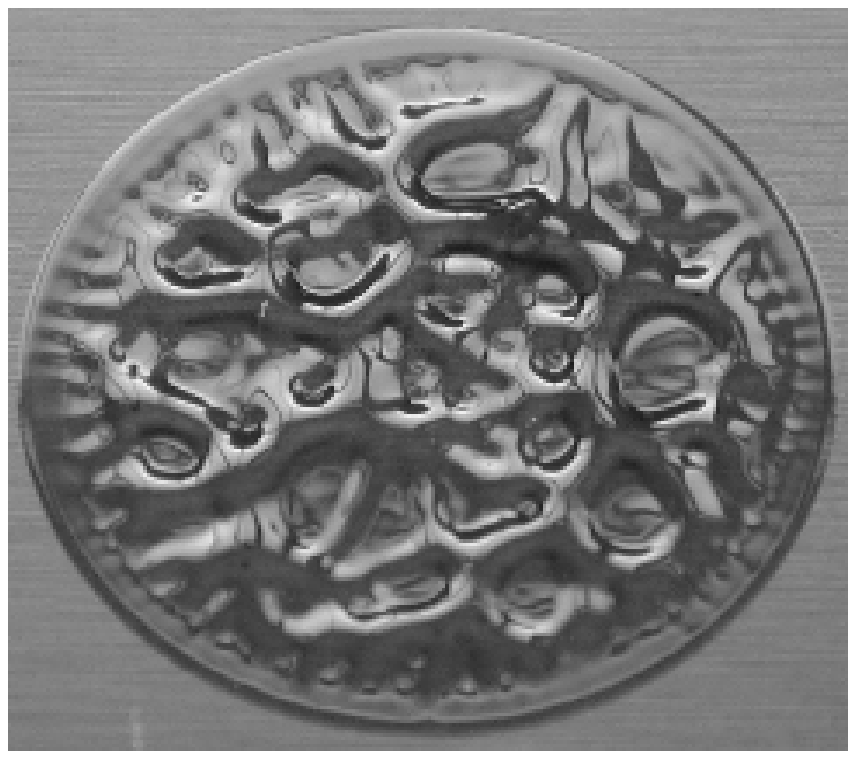}
\end{minipage}\hfill
\begin{minipage}{0.33\linewidth}
\center\includegraphics[width=\linewidth,height=0.75\linewidth]{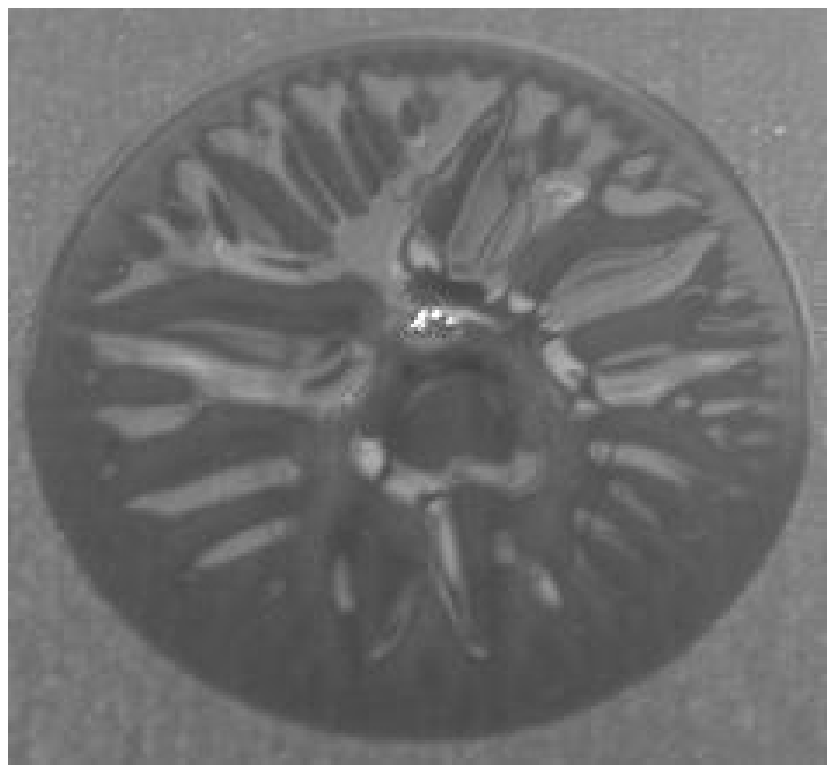}

\vspace{0.5\baselineskip}

\includegraphics[width=\linewidth,height=0.75\linewidth]{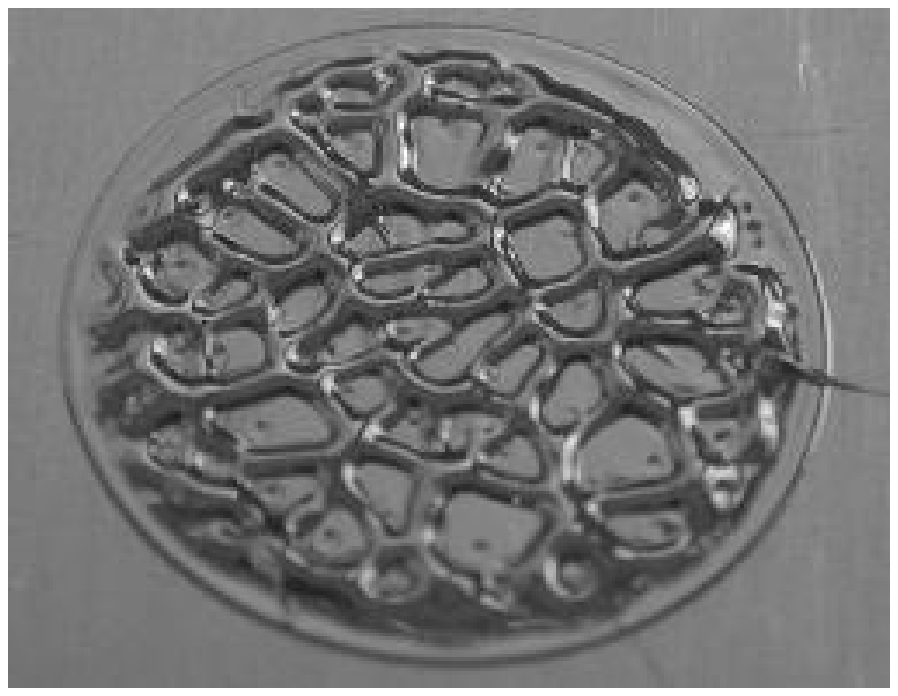}
\end{minipage}
\caption{Patterns left on the bottom plate immediately after full
separation was achieved (increasing velocities from left to right and
from top to bottom; top right is $V=V_c$). Note the arborescent
pattern~\protect\cite{TARAFDAR} at low velocities and the cellular
pattern for $V>V_c$.}
\label{zwick}
\end{figure}

Force-displacement curves obtained at different 
traction velocities 
%ranging from $V=1\mu{\rm m/s}$ to $V=1{\rm mm/s}$ 
are shown in Fig.~\ref{tous-regimes}. 
Two different types of behaviors for the force decrease are clearly
observed. 
The force decrease is smooth at low velocities. 
By contrast, a force plateau immediately followed by a sharp drop 
appears at high velocities, 
a feature strikingly reminiscent of adhesives.
Moreover, a weak noise, resembling the ``pop'' produced by the opening
of a bottle of wine, is heard distinctly in this case. The
transition between these two behaviors can be assigned to the first
curve that displays an inflection point, namely $V_c\simeq 15\mu{\rm
m/s}$.
Fig.~\ref{bepego} shows pictures of the glass--liquid interface taken
during the traction and the associated force curves (lin-log
representation), for three different velocities.
At low velocity (Fig.~\ref{bepego}a, 
$V<V_c$), Saffman--Taylor instabilities develop and
the resulting progressive finger growth corresponds to the smooth
decrease of the force.
At high velocity (Fig.~\ref{bepego}c, $V>V_c$), cavitation is observed
together with the force plateau. More precisely, the pictures show
that numerous bubbles appear just before the force peak and that they
grow and develop in the course of the plateau.  Thus, in purely
viscous liquids and in adhesives alike, the existence of a plateau is
associated to cavitation~\cite{CRETON}.
At $V\simeq V_c$ (Fig.~\ref{bepego}b), one clearly sees the occurence
of both fingering and cavitation together, with the growth of a single
bubble and its subsequent collapse.  This is confirmed by pictures of
the sample taken immediately after full separation (Fig.~\ref{zwick}):
only above $V_c\simeq 15\mu{\rm m/s}$ (Fig.~\ref{zwick}, top right) do
cells appear in the center of the sample, suggesting that cavitation
has taken place.

\vspace{\baselineskip}

From these observations, we can suggest an interpretation of what
happens during the traction over the whole velocity range. The general
idea is that as the plates are being pulled apart, volume conservation
{\em a priori} implies the existence of a radial, convergent flow from
the edge of the sample towards the center, which tends to relieve the
stress. 
Fingering instabilities and cavitation compete in further relieving
the applied stress.  The reason why fingering appears at low
velocities while cavitation takes over at high velocities is the lower
threshold force for fingering
%than for cavitation 
and the higher bubble growth rate for cavitation~\cite{AILLEURS,HETEROGENEOUS}.

Let us now focus on high velocities and in particular on the
unexpected features of the force response: the plateau and the
subsequent force drop. As described above, the force plateau
corresponds to the bubble growth. The explanation goes as
follows. The pressure recorded on the plate is the difference between
the atmospheric pressure and the pressure in the sample.
In the presence of bubbles, the pressure in the sample is the sum of a
contribution from the pressure inside the bubbles and a contribution
(which is small in the present situation) from the flow in the liquid
that surrounds the growing bubbles~\cite{DIV}.  The pressure inside
the bubbles results from two possible sources. {\em(i)} If gas was
present initially in the bubbles, the tremendous volume increase has
made its contribution to the pressure to practically vanish. {\em(ii)}
If the liquid is somewhat volatile or contains a volatile component,
the vapor phase may contribute a pressure whose value cannot exceed
the saturating vapor pressure $p_{sat}$, in which case, the plateau
corresponds to a liquid--vapor phase transition.
In both cases, the bubble pressure contribution is essentially
constant.

At some point of the bubble growth, a few liquid walls between bubbles
break, thus allowing air from the outside to suddenly rush into the
sample from the edge.
This produces the observed abrupt drop of the force, since the
interior of the former bubbles now communicates with the outside air,
thus relieving the pressure difference. This is also responsible for
the ``pop'' which can be heard distinctly as mentioned above.
The total pressure on the plate is thus
expected to drop by $p_{atm}-p_{sat}$ at the end of the force plateau.
The curves in Fig.~\ref{tous-regimes} show that the force drop at the
end of the plateau is on the order of $5\,{\rm N}$, for all
separation rates $V$ in the high velocity regime ($V\geq V_c$). Since
the sample diameter is $9\,{\rm mm}$, this corresponds to a pressure
of about $1\,{\rm atm}$.

\begin{figure}
\center
\includegraphics[width=0.9\linewidth]{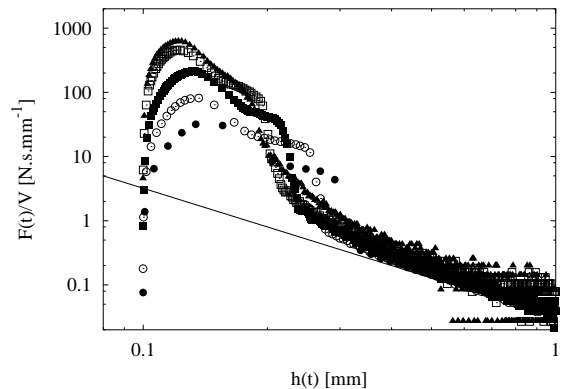}
\caption{Force (normalized by the traction velocity) versus plate
displacement for $V$ ranging from 
$V=V_c=15\mu{\rm m/s}$ ($\blacktriangle$) 
to $V=1000\mu{\rm m/s}$ ($\bullet$). 
The straight line corresponds to
equation~(\ref{fh2}) without any adjustable parameters.}
\label{regime-rapide}
\end{figure}

The pressure difference is now relieved, and there remains the sole
contribution from the liquid walls.  From basic
hydrodynamics~\cite{LANDAU}, the flow is elongational and the force
can be expressed as:
\begin{equation} 
\label{fh2} 
F(t)=\frac{4\eta\,\Omega\,V}{h^2(t)} 
\end{equation} 
where $\Omega$ is the total volume of the walls and $h(t)$ the plate
separation at time $t$.
Fig.~\ref{regime-rapide}, drawn without any adjustable parameters,
shows that the estimate of the force is quite good over a 
velocity range of almost two orders of magnitude.

\begin{figure}
\center
\includegraphics[width=0.9\linewidth]{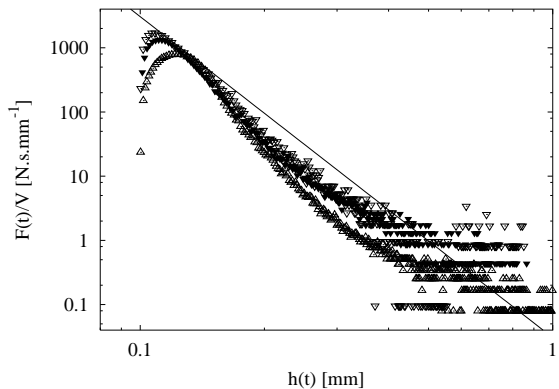}
\caption{Force (normalized by the traction velocity) versus plate
displacement for $V<V_c$. The straight line corresponds to
equation~(\ref{fh5}) without any adjustable parameters.}
\label{regime-lent}
\end{figure}

Let us now turn to the low velocity regime. There is no cavitation.
If we neglect fingering instabilities, the flow at early times is
radial and convergent, 
and locally parabolic (Poiseuille flow)~\cite{LANDAU}. 
The force exerted on the plates is then:
\begin{equation} 
\label{fh5} 
F(t)=\frac{3}{2\pi}\frac{\eta\,\Omega^2\,V}{h^5(t)} 
\end{equation} 
Fig.~\ref{regime-lent} shows that despite the fingering, this law is
quantitatively verified at early times for low separation rates, 
once again without any adjustable parameters. 
At slightly later times, {\it i.e.}, as fingers
have sufficiently developed, thus relieving the stress, the real force
decreases somewhat more rapidly.
At even later times, we might expect equation~(\ref{fh2}) to be valid
($F\propto h^{-2}V$). However, at such low velocities, the
corresponding force lies within the experimental resolution and is,
furthermore, masked by capillarity~\cite{AILLEURS}.

\vspace{\baselineskip}

Let us now compare the behavior of purely viscous liquids to
adhesives.  The mechanisms related to the plateau and the subsequent
force drop are similar~\cite{CRETON,CREVOISIER}, even though the
plateau is generally much higher and much longer in adhesives.  The
height of the plateau is explained by the contribution from the
viscoelastic flow, which is not negligible any more 
compared to atmospheric pressure.  
The walls are also more
resistant to air penetration, explaining the length of the plateau.
As for the force drop, all these observations and interpretations 
lead us to believe that whatever the
material, a sudden air penetration in the sample produces a pressure
drop of $1\,{\rm atm}$.

\vspace{\baselineskip}

By conducting traction experiments on confined, purely viscous
liquids, we have observed and interpreted new behaviors, namely
cavitation and its competition with fingering instabilities. A
remarkable feature of the force response is the appearance of a
plateau at high velocities as observed in adhesives. It can be
interpreted in terms of the liquid being eventually full of almost
empty bubbles. This shows that the bubbles, the force peak and the
plateau are far from specific to adhesive materials: only the relative
importance of two contributions to the plateau force differs. The
appearance of a force plateau for simple liquids is thus interesting
not only {\em per se}, but also in that it brings new elements to
understand what happens in adhesive materials.

\vspace{\baselineskip}

\begin{acknowledgments}
%\section*{Acknowledgments}
We gratefully acknowledge the financial help from the conseil
r\'egional d'Aquitaine, contract number 2000--0201010.
This work would not have been possible without the CRPP instrument
building team who made the apparatus used for our experiments. We are
grateful to Didier Roux for numerous and stimulating discussions.
\end{acknowledgments}


\begin{thebibliography}{99}
\bibitem{WHYTOP}
Our choice to observe the liquid--indenter interface 
is motivated by the fact that 
phenomena leading to rupture 
are most usually triggered there,
as this interface is weaker~\cite{BUB99}.
\bibitem{HETEROGENEOUS} 
Cavitation is clearly heterogeneous, 
as the pressures involved do not exceed a few atmospheres. 
Nuclei can be solid particles in the bulk of the sample. 
They can also be micron sized air bubbles 
trapped at the interface between the fluid and the probe 
during contact formation, as discussed in ref~\cite{BUB99}.
\bibitem{DIV}
This contribution can
be shown to decrease as the bubbles grow, and we believe that it is
mainly responsible for the slight force decrease during the plateau 
and for the plateau shift with traction velocity.
\end{thebibliography}
\end{document}